\documentclass[aps,prx,twocolumn,amsmath,amssymb,superscriptaddress,longbibliography]{revtex4-2}
\usepackage[english]{babel}
\usepackage{xcolor}
\usepackage{amssymb}
\usepackage{dcolumn}
\usepackage{bm}
\usepackage{graphicx}
\usepackage{amsmath}
\usepackage{braket}

\allowdisplaybreaks[4]   

\usepackage{graphicx}        

\graphicspath{{pict/}{}}

\usepackage[normalem]{ulem}
\usepackage{dcolumn}
\usepackage{bm}
\usepackage[pdfstartview=FitH, CJKbookmarks=true, bookmarksnumbered=true, bookmarksopen=true, colorlinks=true, pdfborder=001, citecolor=blue, linkcolor=blue, urlcolor=blue, linktocpage=true] {hyperref}

\begin{document}

\title{Non-Abelian interference of topological edge states}

\author{Shi Hu}
\affiliation{School of Optoelectronic Engineering, Guangdong Polytechnic Normal University, Guangzhou 510665, China}

\author{Meiqing Hu}
\affiliation{School of Optoelectronic Engineering, Guangdong Polytechnic Normal University, Guangzhou 510665, China}

\author{Zhoutao Lei}
\email{leizht3@mail2.sysu.edu.cn}
\affiliation{Guangdong Provincial Key Laboratory of Quantum Metrology and Sensing $\&$ School of Physics and Astronomy, Sun Yat-Sen University (Zhuhai Campus), Zhuhai 519082, China}
\date{\today}

\begin{abstract}
Topological boundary states exhibit distinctive properties, including unidirectional propagation and noise robustness, which hold significant potential for advancing the performance of quantum science and technology. 
Here, we demonstrate the implementation of non-Abelian quantum interference and entanglement generation, protected by dual symmetries (time-independent inversion and time-dependent interchain), in coupled Su-Schrieffer-Heeger  chains.
Specifically, in a multi-chain system, we first achieve tunable topological transfer of a single particle, where the destination chain is selected by the permutation sequence.
We then extend this to two particles, observing a non-Abelian Hong-Ou-Mandel interference that generates spatially entangled NOON states whose properties are dictated by the permutation sequence.
Our work establishes an alternative pathway for exploring non-Abelian topology applied to quantum science and technology, enabled by the unique protection of time-dependent symmetry.
\end{abstract}
\maketitle

\section{Introduction}\label{Sec1}
Topological quantum matters have emerged as a pivotal impetus for fundamental research, with implications spanning diverse fields from condensed-matter physics \cite{MZHasanRMP2010,XLQiRMP2011} to quantum information processing \cite{CNayakRMP2008,RMLutchynPRL2010}.
Typical paradigms of topological states encompass the Chern insulators \cite{KlitzingPRL1980,ThoulessPRL1982}, topological insulators with $Z_2$ topological invariants \cite{KanePRL2005,KaneGraphenePRL2005,BernevigPRL2006}, topological superconductor \cite{MourikScience2012,DasNP2012}, and topological semimetals \cite{XuScience2015,SongPRX2019}.
One of the most striking features of topological systems is the existence of topologically protected boundary states, dictated by the bulk-boundary correspondence \cite{YHatsugaiPRL1993,EssinPRB2011}, which exhibit unique properties such as robustness anchored in bulk topological invariants \cite{RyuNJP2010,CKChiuRMP2016} and unidirectional propagation immune to backscattering \cite{ZWangNature2009,JSeoNature2010}.
The considerable diversity of topological boundary states, discovered in systems of different dimensions and symmetries, has served as a source for numerous studies across quantum science and technology, particularly in quantum state transfer \cite{YEKrausPRL2012,NLangQI2017,CDlaskaQST2017,FMeiPRA2018,SLonghiPRB2019,NEPalaiodimopoulosPRA2021,LHuangPRA2022,CWangPRA2022}, quantum gates \cite{PBorossPRB2019,MNarozniakPRB2021}, and advanced quantum devices \cite{RHammerPRB2013,XSWangPRB2017,LQiPRB2021,LQiPRA2023}.

In parallel to these developments, the role of fundamental quantum effects in such systems has garnered significant attention.
Specifically, quantum entanglement and interference play a crucial role in the three major fields of quantum science and technology, including quantum communication, quantum computation, and quantum metrology \cite{AEkertRMP1996,JWPanRMP2012,LPezzeRMP2018}.
Topological systems have been shown to facilitate the topological protection of quantum entanglement \cite{MCRechtsmanOptica2016,MWangNanophotonics2019,KMonkmanPRR2020,JXHanPRA2021}, highlighting their potential for stabilizing quantum effects against external perturbations.
For instance, the well-known Hong-Ou-Mandel (HOM) interference \cite{CKHongPRL1987} of topological edge states has recently been observed in experiments on integrated photonic circuits \cite{JLTambascoSA2018}.
Moreover, beyond experimental observations, several theoretical proposals have also been put forward to achieve interference of topological states, generate spatially entangled two-particle NOON states, and realize nonadiabatic topological transport protocols \cite{SHuPRA2020,SHuPRA2024,SHuPRA2025}.
Notably, the frontier of this exploration is now extending into the realm of non-Abelian topology \cite{KitaevAP2003,AliceaNP2011,Sarmanpj2011,WuScience2019,BouhonNP2020,BouhonPRB2020,BouhonPRB2021,GuoNature2021,BroscoPRA2021,SunNP2022,PengNC2022,LiNC2023,PengNC2024,SunPRL2024,BreachPRL2024,ChenPRR2025}. 
In such systems, their non-commutative properties enriching the landscape of quantum phenomena.
Despite this intriguing potential, research on quantum entanglement and interference mediated specifically by non-Abelian topological edge states remains largely unexplored.

To address this gap, and motivated by recent advances in topological insulators with exotic gauge fields \cite{YXZhaoNC2022,XuePRL2022,LiPRL2022,LongPRL2024} and quantum anomalous semimetals \cite{Funpj2022}, we introduce a periodically modulated one-dimensional multi-chain system as a platform for non-Abelian quantum interference and entanglement generation via permutation of topological edge states. 
The non-Abelian characteristics of the platform are protected by a time-dependent interchain symmetry in two crucial aspects: First, it enables the formation of multiple edge-state branches with distinct symmetry eigenvalues, whose temporal permutation induces branch switching, which directly exhibits the non-Abelian properties. Second, it ensures the robust adiabatic evolution of states along a single branch, even when the energy gap to other branches closes transiently.
Additionally, the dynamics are also governed by a time-independent inversion symmetry that protects the adiabatic evolution of parity-distinct edge states, with a relative phase accumulating to determine the arbitrary partition of particle transfer at the chain ends.
Specifically, when each edge-state branch is initially localized on a separate chain (with uncoupled initial conditions), a particle injected into one end of a chain (as a superposition of parity-opposite states in its associated branch) can be transferred to the two ends of another chain at a tunable ratio (0 to 1).
The non-Abelian character of this transport is evidenced by the fact that the choice of the recipient chain is determined by the sequence of permutation, while the final partition within that chain is controlled by the accumulated dynamical phase.
When extended to two particles, our system allows for the observation of a non-Abelian HOM interference, a process in which the permutation sequence dictates the generated entangled state.

This article is organized as follows.
In Sec.~\ref{Sec2}, we discuss the permutation of two pairs of topological edge states protected by the time-dependent interchain symmetry in a double-coupled SSH chain.
In Sec.~\ref{Sec3}, we extend our analysis to a triple-coupled SSH chain system and achieve topological transport, HOM interference, and entanglement generation based on the non-Abelian permutation of three pairs of edge states.
Finally, a summary and discussion are provided in Sec.~\ref{Sec4}.

\section{Permutation of two pairs of topological edge states in a double-coupled chain}\label{Sec2}
\begin{figure}[!htp]
\includegraphics[width=1\columnwidth]{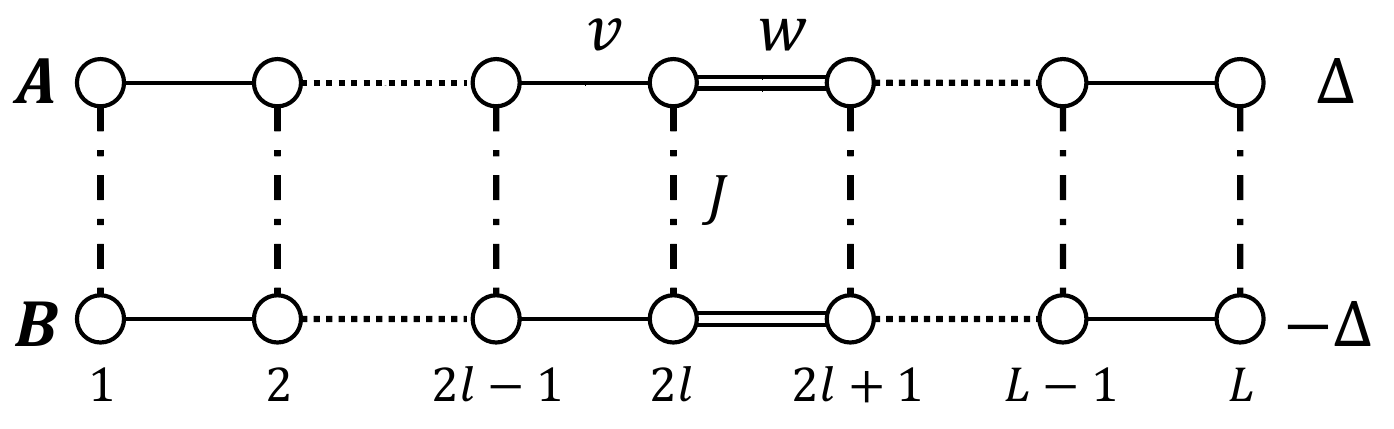}
\caption{\label{Model1}Schematic diagram of double-coupled SSH chains $A$ and $B$. $v$ and $w$ denote the alternate intrachain hopping, $J$ represent the interchain hopping, and $2\Delta$ are on-site energy offsets.}
\end{figure}

In this section, we focus on the permutation of two pairs of topological edge states in a double-chain system.
While this system does not exhibit non-Abelian characteristics, its relative simplicity makes it an ideal starting point for introducing the fundamental mechanism.
Specifically, we consider double-coupled Su-Schrieffer-Heeger (SSH) chains, labeled $A$ and $B$, with opposite global on-site energies. The system is governed by the Hamiltonian (with $\hbar=1$):
\begin{eqnarray}\label{HamHD}
&&\hat{H}_{D}
=\hat{H}_{D\text{-}{\rm Inter}}+\hat{H}_{D\text{-}{\rm SSH}},\cr\cr
&&\hat{H}_{D\text{-}{\rm Inter}}
=\sum_{l}\hat{\bm{b}}_{l}^{\dag}{\bm h}_2\hat{\bm{b}}_{l},
~~{\bm h}_2=\left[
\begin{array}{ll}
\Delta~~~~~
J\\
J~~
-\Delta
\end{array}\right],
\cr\cr
&&\hat{H}_{D\text{-}{\rm SSH}}=\sum_{l}
v\hat{\bm{b}}_{2l-1}^{\dag}{\hat{\bm{b}}_{2l}}
+w\hat{\bm{b}}_{2l}^{\dag}{\hat{\bm{b}}_{2l+1}}
+{\rm H.c.},
\end{eqnarray}
where $\hat{\bm{b}}_{l}^{\dag} = (\hat{b}_{Al}^{\dag}, \hat{b}_{Bl}^{\dag})$ and $\hat{\bm{b}}_{l} = (\hat{b}_{Al}, \hat{b}_{Bl})^T$ with $\hat{b}_{Al}^{\dag}$ ($\hat{b}_{Bl}^{\dag}$) and $\hat{b}_{Al}$ ($\hat{b}_{Bl}$) being the creation and annihilation operators acting on lattice site $l$ of SSH Chain $A$ ($B$), respectively.
Furthermore, the intrachain hopping amplitudes are set as $v = v_0|\sin(\omega t)|$ and $w = 1$ (taken as the energy unit), while the interchain hopping amplitude and the on-site energy offset are $J = \kappa\sin(2\omega t)$ and $\Delta = \kappa[1+\cos(2\omega t)]$, respectively.
In this work, we focus on the topological phase with 
$|v|<|w|$, where topological edge states are localized at the two ends of the chains, with a decay index of $(v/w)^2$ \cite{SHuPRA2024}.
With these parameters, this time-dependent Hamiltonian has a period $T = \pi/\omega$.
An illustrative plot for this double-chain system with $2L$ lattice sites is provided in Fig.~\ref{Model1}.

Similar to the original (single-)SSH chain system, this system preserves  inversion symmetry, which is defined as $\hat{\mathcal{P}}: \hat{\bm{b}}_{l} \rightarrow \hat{\bm{b}}_{L+1-l}$.
Furthermore, beyond the single-chain system, this system also possesses  time-dependent interchain symmetry, which is defined as $\hat{\mathcal{S}}_{D}(t): \hat{\bm{b}}_{l} \rightarrow {\bm s}_{D}(t)\hat{\bm{b}}_{l}$ with the symmetry matrix       
\begin{eqnarray}\label{SD}
{\bm s}_{D}(t)=e^{-i\omega t}
\left[
\begin{array}{ll}
\cos(\omega t)~~~~
\sin(\omega t)  \\
\sin(\omega t)~~
-\cos(\omega t)
\end{array}\right].
\end{eqnarray}
Clearly, this time-dependent interchain symmetry shares the same period as the Hamiltonian.
Since these two symmetry operators commute, i.e., $\hat{\mathcal{P}}\hat{\mathcal{S}}_{D}(t) = \hat{\mathcal{S}}_{D}(t)\hat{\mathcal{P}}$, the eigenstates of the Hamiltonian $\hat{H}_D$ can simultaneously serve as their common eigenstates. 
Specifically, for the single-particle case whose Hilbert space can be spanned by $|l_A\rangle=\hat{b}_{Al}^{\dag}|V\rangle$, $|l_B\rangle=\hat{b}_{Bl}^{\dag}|V\rangle$ with $|V\rangle$ denoting the vacuum state, the symmetry operator $\hat{\mathcal{P}}$ and $\hat{\mathcal{S}}_{D}(t)$ can be represented by ${\bm P}$ and ${\bm S}_{D}(t)$, respectively, with 
\begin{eqnarray}\label{Parity}
{\bm P}&=&\sum_l^L\sum_{\alpha=A,B}|l_{\alpha}\rangle\langle (L+1-l)_{\alpha}|,\cr\cr
{\bm S}_{D}(t)
&=&\sum_l^L\sum_{\alpha=A,B}\sum_{\beta=A,B}[{\bm s}_{D}(t)]_{\alpha\beta}|l_{\alpha}\rangle\langle l_{\beta}|.
\end{eqnarray}
Both $\bm{P}$ and $\bm{S}_D(t)$ possess two eigenvalues: $\bm{P}$ has $\pm1$ (corresponding to odd/even parity), and $\bm{S}_D(t)$ has $S_{D,\text{I}}(t) = -S_{D,\text{II}}(t) = e^{-i\omega t}$.
We note that the eigenvalues of $\bm{S}_D(t)$ exhibit a swapping property: $S_{D,\text{I}/\text{II}}(T) = S_{D,\text{II}/\text{I}}(0)$, with their associated eigenstates behaving identically.
This induces intriguing phenomena in the system's static energy spectrum and dynamical evolution, as discussed below.  
It is worth noting that when constructing the double-coupled SSH chains, we first adopt a well-studied non-Hermitian Hamiltonian of a two-level system with states featuring braiding characteristics \cite{HuPRL2021,HuPRR2022,GuoPRL2023,WangNanotechnology2025} as the symmetry operator of our system.
We then reverse-engineer the interchain term $h_2$ based on the commutation relation, from which the Hamiltonian $\hat{H}_{D}$ (presented in Eq.~\eqref{HamHD}) is derived.
Notably, this method is equally applicable to the triple-coupled SSH chains discussed subsequently.

\subsection{Twisted energy spectrum and permutation of two pairs of edge states}\label{Sec21}
\begin{figure}[!htp]
\includegraphics[width=1\columnwidth]{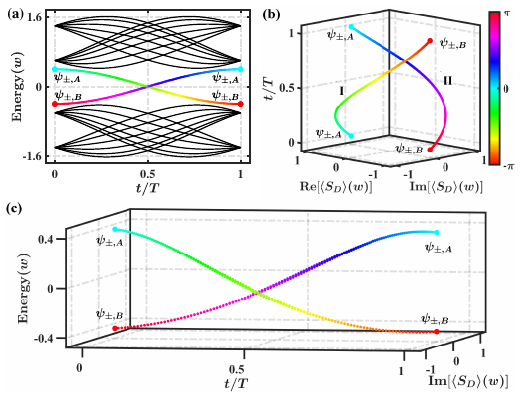}
\caption{\label{ABraiding}permutation of two pairs of topological edge states.
(a) Energy spectrum of the double-coupled SSH chains. Black solid lines represent the bulk states, and colored lines represent pairs of edge states. 
(b) Two branches of the expectation value $\langle\bm{S}_D\rangle$.
(c) Abelian braiding processes of two pairs of edge states in the ($t$, $\text{Im}[\langle\bm{S}_D\rangle]$, $E$) space, corresponding to the braiding operators $\tau^{-1}_{1}$.
In all panels, the color denotes the argument of the expectation value $\langle\bm{S}_D\rangle$.
The parameters set as $L=14$, $w=1$, $v_{0}=0.5$, and $\kappa=0.2$.}
\end{figure}
We first analyze the static energy spectrum constrained by inversion symmetry and time-dependent interchain symmetry.
As depicted in Fig.~\ref{ABraiding}(a), four edge states emerge in the gap.
Explicitly, these edge states are also common eigenstates of $\bm{P}$ and $\bm{S}_D(t)$, and thus can be indexed as $|\psi_{\pm,\text{I}/\text{II}}\rangle$ with energies $\varepsilon_{\pm,\text{I}/\text{II}}$ according to the eigenvalues of the symmetry operators.
Specifically, edge states sharing the same eigenvalues of $\bm{S}_D(t)$ exhibit a small yet equal energy difference, where $\varepsilon_{+,\text{I}} - \varepsilon_{-,\text{I}} = \varepsilon_{+,\text{II}} - \varepsilon_{-,\text{II}}$, and this difference even vanishes at $t=0$ and $t=T$.
Therefore, these four eigenstates form two branches.
To better characterize the properties of these edge states, we further color-code them in the spectrum of Fig.~\ref{ABraiding}(a) according to the argument of their expectation value with respect to $\bm{S}_D(t)$.
As observed here, for Branch I, the argument varies from $0$ to $-\pi$, while for Branch II, it varies from $\pi$ to $0$, this behavior is consistent with the eigenvalues of $\bm{S}_D(t)$, i.e., $\pm e^{-i\omega t}$.
More importantly, corresponding to the swapping property of the eigenvalues of $\bm{S}_D(t)$ [i.e., $S_{D,\text{I}/\text{II}}(T) = S_{D,\text{II}/\text{I}}(0)$], the energies of these edge states also exhibit this property: $\varepsilon_{\pm,\text{I}/\text{II}}(T)= \varepsilon_{\pm,\text{II}/\text{I}}(0)$. 
This behavior manifests as a twisted structure of the two branches in the spectrum [Fig.~\ref{ABraiding}(a)].

More intuitively, Fig.~\ref{ABraiding}(b) visualizes the real and imaginary parts of the edge states' expectation value with respect to $\bm{S}_D(t)$ (denoted as $\langle\bm{S}_D\rangle=\langle\psi|\bm{S}_D(t)|\psi\rangle$), and Fig.~\ref{ABraiding}(c) depicts the variation of the branches formed by the imaginary parts of $\langle\bm{S}_D\rangle$ and the energies.
Here, the braiding behavior of the edge states is clearly visible.
Specifically, within the single period depicted in Fig.~\ref{ABraiding}(c), the braid operation $\tau_{1}^{-1}$ occurs, where we define $\tau_n$ ($\tau_n^{-1}$) as the braid operation in which the $n$-th strand crosses under (above) the $(n+1)$-th strand.
This induces the aforementioned switching of the edge states' energy and their association with $\bm{S}_D(t)$.
The braiding processes in the ($t$, $\text{Im}[\langle\bm{S}_D\rangle]$, $E$) parameter space result in the swap of two pairs of topological edge states, as discussed in the following.
It is worth noting that the phase factor in Eq.~\eqref{SD} can be taken as $\mathrm{e}^{\mathrm{i}(2n+1)\omega t}$ with $n\in\mathbb{Z}$ under the condition that the time-dependent interchain symmetry operator $\hat{\mathcal{S}}_{D}(t)$ and the Hamiltonian $\hat{H}_{D}$ share the same period, and different values of $n$ may reverse the braiding operator (i.e., $\tau_{1}\leftrightarrow\tau_{1}^{-1}$).
However, both $\tau_{1}$ and $\tau_{1}^{-1}$ can be mapped to the permutation operation $\mathcal{M}_{1}$, where $\mathcal{M}_{n}$ is defined as the permutation operation that swaps the $n$-th pair of topological edge states with the $(n+1)$-th pair.
It is convenient to describe the swap of these two pairs of edge states using the permutation operation.

Given the periodicity of the Hamiltonian and the symmetry operator $\bm{S}_D(t)$, i.e., $\hat{H}_D(T) = \hat{H}_D(0)$ and $\bm{S}_D(T) = \bm{S}_D(0)$, we ultimately conclude that the two pairs of edge states swap after one period, following $|\psi_{\pm,\text{I}/\text{II}}(T)\rangle = |\psi_{\pm,\text{II}/\text{I}}(0)\rangle$ described by the permutation operation $\mathcal{M}_{1}$.
This phenomenon is more pronounced under the parameter variation considered herein, which has been presented below Eq.~\eqref{HamHD}: at the initial and final moments of the period, the intrachain SSH segment of the system is in the fully dimerized limit ($v=0$), and the interchain coupling is decoupled ($J=0$).
Such a configuration implies that at these two instants, the eigenedge states are fully localized odd-even superposition states at the two ends of a single chain. 
By further incorporating information on the energy and the expectation values of the symmetry operator, we obtain the following expressions:
\begin{eqnarray}
\label{PsiAB}
|\psi_{\pm,\text{I}}(0)\rangle&=&|\psi_{\pm,\text{II}}(T)\rangle = |\psi_{\pm,A}\rangle,\cr\cr
|\psi_{\pm,\text{II}}(0)\rangle&=&|\psi_{\pm,\text{I}}(T)\rangle = |\psi_{\pm,B}\rangle,\cr\cr
|\psi_{\pm,A/B}\rangle&=&\frac{|1_{A/B}\rangle \pm |L_{A/B}\rangle}{\sqrt{2}},
\end{eqnarray}
indicating that the swapping of the edge states can be straightforwardly confirmed by the exchange of their positions, i.e., between Chain A and Chain B.

\subsection{Symmetry-protected adiabatic evolution}\label{Sec22}
We now turn to the adiabatic evolution of the edge states under the protecting of inversion symmetry and time-dependent interchain symmetry.
Firstly, the parity of the states is conserved despite their extremely small energy difference during dynamical evolution, owing to inversion symmetry.
Similarly, the interchain symmetry can also protect the adiabatic evolution of the edge states.
This protection is manifested in the independent evolution of edge states in the two branches. 
Even though these two branches exhibit energy crossing, this feature would imply infinite adiabatic velocity in the absence of symmetry protection.
We note that, in contrast to time-independent symmetries (such as inversion symmetry), this adiabatic evolution under the protection of time-dependent symmetry still requires a sufficient long (yet not infinite) evolution time, as discussed in the Appendix.
Thereby, by combining them with the permutation properties (another effect brought by the time-dependent interchain symmetry), the edge states can propagate from the left end to the right end of the system during adiabatic evolution, and also transition from one chain to the other.

\begin{figure}[!htp]
\includegraphics[width=1\columnwidth]{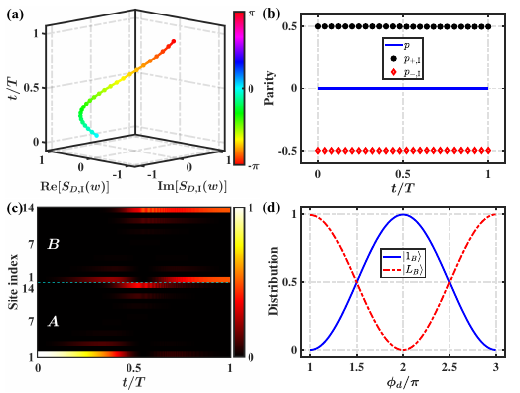}
\caption{\label{Evolution}Symmetry-protected adiabatic evolution. 
(a) Expectation values $\langle\bm{S}_D\rangle$ (colored dots) and eigenvalues $S_{D,\text{I}}$ (colored lines) during adiabatic evolution. The color indicates the argument of $\langle\bm{S}_D\rangle$ or $S_{D,\text{I}}$.
(b) Conservation of parity during adiabatic evolution.
(c) Density distribution $\langle\hat{n}_{j}\rangle$ as a function of evolution time $t$.
(d) Distribution of the final state at the two ends ($|1_{B}\rangle$ and $|L_{B}\rangle$) of chain $B$ versus $\phi_{d}\in[\pi,3\pi]$.
The parameters set as $L=14$, $w=1$, $v_{0}=0.5$, $\kappa=0.2$.
For panels (a), (b), and (c), $T=1332$ (corresponding to $\phi_{d}=1.5\pi$).}
\end{figure}

More specifically, we consider the adiabatic evolution with an initial state at the left end of Chain A, which can be expressed as a superposition of edge states in Branch I:
\begin{eqnarray}\label{evot}
|\varphi(t)\rangle &=& \mathcal{T}e^{-i\int_{0}^{t}\hat{H}_D(\tau) d\tau}|\varphi(0)\rangle,  \cr\cr
|\varphi(0)\rangle &=& |1_A\rangle = \frac{|\psi_{+,\text{I}}(0)\rangle + |\psi_{-,\text{I}}(0)\rangle}{\sqrt{2}},
\end{eqnarray}
with $\mathcal{T}$ being the time-ordering operator.
We present the expectation value $\langle\bm{S}_D\rangle$ during the evolution with a sufficiently slow velocity (here, $T=1332$ in units of $1/w$) in Fig.~\ref{Evolution}(a), which are in good agreement with the Branch I eigenvalues $S_{D,\text{I}}(t)$, implying that $|\varphi(t)\rangle$ remains within this branch.
Furthermore, we also present the parity $p(t)$ and projected parity $p_{\pm,\text{I}}(t)$ on Branch I during the evolution in Fig.~\ref{Evolution}(b), with these quantities defined as follows:
\begin{eqnarray}\label{parity}
p(t)&=&\langle\varphi(t)|\bm{P}|\varphi(t)\rangle^{2},\cr\cr
p_{+,\text{I}}(t)&=&|\langle\psi_{+,\text{I}}(t)|\varphi(t)\rangle|^{2},\cr\cr 
p_{-,\text{I}}(t)&=&-|\langle\psi_{-,\text{I}}(t)|\varphi(t)\rangle|^{2}. 
\end{eqnarray}
As illustrated in Fig.~\ref{Evolution}(b), the parity $p(t)=0$, and the projected parities $p_{+,\text{I}}(t)=0.5$ and $p_{-,\text{I}}(t)=-0.5$.
It verifies that, due to inversion symmetry, the components with different parities in the initial state (here, $|1_A\rangle$) evolve separately, as mentioned earlier.
Thereby, after a period of adiabatic evolution, the final state can be expressed up to an overall phase as:
\begin{eqnarray}\label{evoend}
|\varphi(T)\rangle &=& \frac{e^{-i\frac{\phi_d}{2}}|\psi_{+,\text{I}}(T)\rangle + e^{i\frac{\phi_d}{2}}|\psi_{-,\text{I}}(T)\rangle}{\sqrt{2}} \cr\cr
&=& \cos\frac{\phi_d}{2}|1_B\rangle -i\sin\frac{\phi_d}{2}|L_B\rangle,
\end{eqnarray}
where $\phi_d = \int_{0}^{T} \left[ \varepsilon_{+,\text{I}}(t) - \varepsilon_{-,\text{I}}(t) \right] dt$ is the accumulated phase difference.
The second equality here makes use of the relations given in Eq.~\eqref{PsiAB}.
This result reveals that the particle initially injected into the left end of Chain $A$ can be transferred to the two ends of Chain $B$ with a tunable ratio ranging from $0$ to $1$, depending on the relative dynamical phase $\phi_d$.
In Fig.~\ref{Evolution}(c), we present the density distribution $\langle\hat{n}_{j}\rangle=\langle\hat{b}_{j}^{\dag}\hat{b}_{j}\rangle$ as a function of the evolution time $t$.
Here, we choose $T=1332$ and $L=14$, which corresponds to $\phi_{d}=1.5\pi$.
The final density distribution, which is uniformly distributed at the two ends of chain $B$, is in good agreement with the analytical results in Eq.~\eqref{evoend}.
Additionally, the fidelity $F=|\langle\psi|\varphi(T)\rangle|$ between the final state $|\varphi(T)\rangle$ and the target state $|\psi\rangle=-(|1\rangle_{B}+i|L\rangle_{B})/\sqrt{2}$ [predicted by Eq.~\eqref{evoend}] reaches $0.998$.
To further exhibit the effect of the accumulated dynamical phase difference $\phi_{d}$, we plot the distribution of the final state at the two ends ($|1_{B}\rangle$ and $|L_{B}\rangle$) of Chain $B$ versus $\phi_{d}\in[\pi,3\pi]$ in Fig.~\ref{Evolution}(d), where $|1_{B}\rangle$ is represented by a blue solid line and $|L_{B}\rangle$ by a red dashed line. These numerical results are in good agreement with the analytical ones in Eq.~\eqref{BS}.

When the initial state is $|L\rangle_{A}$, we obtain a similar result, both summarized as follows:
\begin{eqnarray}\label{BS}
|1_{A}\rangle &\rightarrow \cos\frac{\phi_{d}}{2}|1_{B}\rangle  -i\sin\frac{\phi_{d}}{2}|L_{B}\rangle, \cr\cr
|L\rangle_{A} &\rightarrow -i\sin\frac{\phi_{d}}{2}|1\rangle_{B} + \cos\frac{\phi_{d}}{2}|L\rangle_{B},
\end{eqnarray}
where an equal overall phase $\phi_o = \frac{1}{2}\int_{0}^{T} \left[ \varepsilon_{+,\text{I}}(t) + \varepsilon_{-,\text{I}}(t) \right]dt$ is neglected.
Evidently, when the initial state is placed at one end of chain $B$, the results are symmetric to those above: the final states appear at the ends of Chain A, and further elaboration is omitted here.

\section{Non-Abelian permutation and Hong-Ou-Mandel interference in a triple-coupled chain}\label{Sec3}
\begin{figure}[!htp]
\includegraphics[width=1\columnwidth]{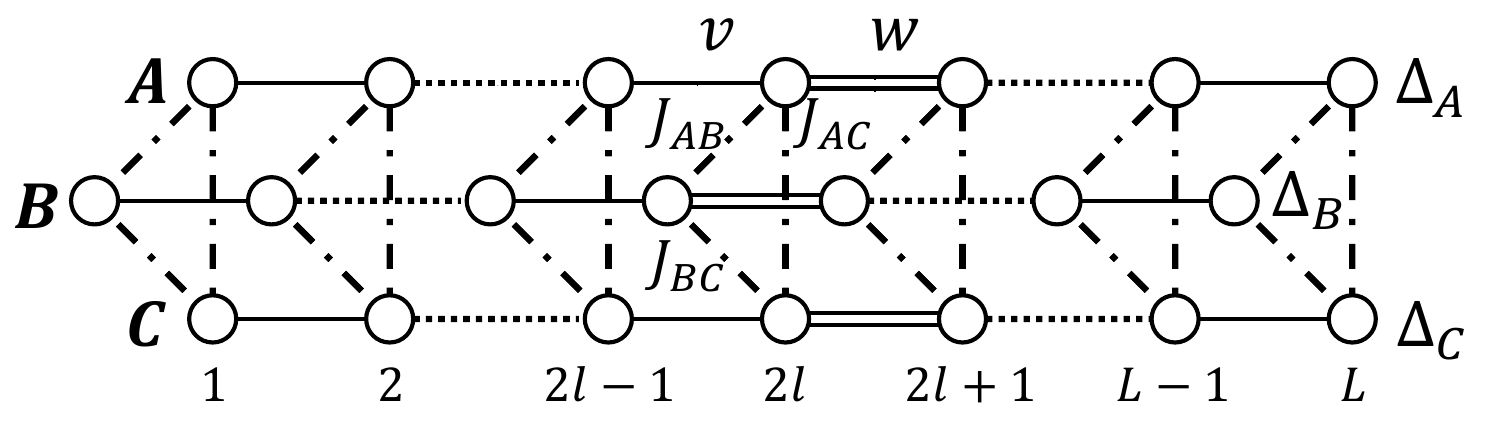}
\caption{\label{Model2}Schematic diagram of triple-coupled SSH chains $A$, $B$ and $C$. $v$ and $w$ denote the alternate intrachain hopping. $J_{AB}$, $J_{AC}$, and $J_{BC}$ represent the interchain hopping amplitudes between chain $A$ and $B$, $A$ and $C$, and $B$ and $C$, respectively, while $\Delta_{A}$, $\Delta_{B}$, and $\Delta_{C}$ denote the on-site energies of chain $A$, $B$, and $C$, respectively.}
\end{figure}
The permutation of two pairs of topological edge states in Sec.~\ref{Sec2} demonstrates the protection of time-dependent interchain symmetry, which we now extend to the case of three pairs described by non-Abelian permutation.
We consider triple-coupled SSH chains denoted as $A$, $B$, and $C$, which are described by the following Hamiltonian (with $\hbar=1$):
\begin{eqnarray}\label{HamHT}
&&\hat{H}_{T}
=\hat{H}_{T\text{-}{\rm Inter}}+\hat{H}_{T\text{-}{\rm SSH}},\cr\cr
&&\hat{H}_{T\text{-}{\rm Inter}}
=\sum_{l}\bm{\hat{b}_{l}^{\dag}}h_3\bm{\hat{b}_{l}},
~~h_3=
\left[
\begin{array}{lll}
\Delta_{A}~~~~J_{AB}~~~~J_{AC}  \\
J_{AB}~~~~\Delta_{B}~~~~J_{BC}  \\ 
J_{AC}~~~~J_{BC}~~~~\Delta_{C}  
\end{array}\right],
\cr\cr\cr
&&\hat{H}_{T\text{-}{\rm SSH}}=\sum_{l}
v\bm{\hat{b}_{2l-1}^{\dag}}\bm{\hat{b}_{2l}}
+w\bm{\hat{b}_{2l}^{\dag}}\bm{\hat{b}_{2l+1}}
+{\rm H.c.},
\end{eqnarray}
where $\bm{\hat{b}_{l}^{\dag}}=(\hat{b}_{Al}^{\dag},\hat{b}_{Bl}^{\dag},\hat{b}_{Cl}^{\dag})$ and $\bm{\hat{b}_{l}}=(\hat{b}_{Al},\hat{b}_{Bl},\hat{b}_{Cl})^{T}$, now including the creation and annihilation operators acting on Chain $C$.
Similar to the double-coupled chains system, the Hamiltonian is also divided into two parts.
For the $\hat{H}_{T\text{-}{\rm SSH}}$ part, the intrachain hopping amplitudes are as same in Eq.~\ref{HamHD}.
For the $\hat{H}_{T\text{-}{\rm Inter}}$ part, as shown in Fig.~\ref{Model2}, $J_{AB}$, $J_{AC}$, and $J_{BC}$ represent the interchain hopping amplitudes between Chain $A$ and $B$, $A$ and $C$, and $B$ and $C$, respectively, while $\Delta_{A}$, $\Delta_{B}$, and $\Delta_{C}$ denote the on-site energies of Chain $A$, $B$, and $C$, respectively.
These parameter variations considered are presented below:
\begin{equation}\label{JAB}
\left\{
\begin{aligned}
J_{AB}&=\eta\left[\frac{1}{3}-\frac{2\sqrt{3}}{9}
\sin(2\omega t+\frac{2\pi}{3}) \right], \\
J_{AC}&=\eta\left[\frac{1}{3}-\frac{2\sqrt{3}}{9}
\sin(2\omega t+\frac{\pi}{3}) \right], \\
J_{BC} &=J_{AB}-J_{AC},
\end{aligned}
\right.
\end{equation}
and
\begin{equation}\label{DeltaAB}
\left\{
\begin{aligned}
\Delta_{A}&=\eta\left[ \frac{2}{3} + \frac{2\sqrt{3}}{9}
\sin(2\omega t+\frac{\pi}{3}) \right], \\
\Delta_{C}&=\eta\left[ -\frac{2}{3} -\frac{2\sqrt{3}}{9}
\sin(2\omega t+\frac{2\pi}{3}) \right],\\
\Delta_{B}&=-\Delta_{A}-\Delta_{C}.
\end{aligned}
\right.
\end{equation}

Analogous to the double-coupled chain system discussed in Sec.~\ref{Sec2}, the triple-coupled system also preserves inversion symmetry $\hat{\mathcal{P}}$ and time-dependent interchain symmetry $\hat{\mathcal{S}}_{T}(t)$.
The inversion symmetry $\hat{\mathcal{P}}$ is straightforwardly extended to this system as $\hat{\mathcal{P}}: \hat{\bm{b}}_{l} \rightarrow \hat{\bm{b}}_{L+1-l}$, which now incorporates the creation and annihilation operators acting on Chain $C$.
The time-dependent interchain symmetry $\hat{\mathcal{S}}_{T}(t)$, described by the map $\hat{\bm{b}}_{l} \rightarrow {\bm s}_{T}(t)\hat{\bm{b}}_{l}$ is associated with the symmetry matrix ${\bm s}_{T}(t)$          
\begin{eqnarray}\label{ST}
\bm{s}_T(t) &=& \frac{1}{3}e^{-2i\omega t}
\begin{bmatrix}
~~~1~~~~~ & e^{\frac{2\pi i}{3}} & e^{\frac{\pi i}{3}} \\
e^{\frac{2\pi i}{3}} & -e^{\frac{\pi i}{3}} & ~~~~-1~~\\
e^{\frac{\pi i}{3}} & ~~~~-1~~~~ & e^{\frac{2\pi i}{3}}
\end{bmatrix} \notag \\\notag \\
&+& \frac{1}{3}
\begin{bmatrix}
~~~2~~~ & -e^{\frac{2\pi i}{3}} & -e^{\frac{\pi i}{3}} \\
-e^{\frac{2\pi i}{3}} & 2e^{\frac{\pi i}{3}} & ~~~~1~~~~ \\
-e^{\frac{\pi i}{3}} & ~~~~~~1~~~~~~ & -2e^{\frac{2\pi i}{3}}
\end{bmatrix},
\end{eqnarray}
where the symmetry operator shares the same period as the Hamiltonian.
Under the single-particle condition, with the Hilbert space spanned by $\{|l_A\rangle, |l_B\rangle, |l_C\rangle\}$, the symmetry operators $\hat{\mathcal{P}}$ and $\hat{\mathcal{S}}_{T}(t)$ can be represented by $\bm{P}$ and $\bm{S}_{T}(t)$, respectively, as follows:
\begin{eqnarray}\label{ReST}
{\bm P}&=&\sum_l^L\sum_{\alpha=A,B,C}|l_{\alpha}\rangle\langle (L+1-l)_{\alpha}|,\cr\cr
{\bm S}_{T}(t)
&=&\sum_l^L\sum_{\alpha=A,B,C}\sum_{\beta=A,B,C}
[{\bm s}_{T}(t)]_{\alpha\beta}|l_{\alpha}\rangle\langle l_{\beta}|.
\end{eqnarray}
The eigenvalues $\pm1$ of $\bm{P}$ can be used to denote the even and odd parity of the edge states, respectively.
What's more, the eigenvalues $S_{T,\text{I}}(t)=e^{2i\pi/3}S_{T,\text{II}}(t)=e^{-2i\pi/3}S_{T,\text{III}}(t)= e^{-2i\omega t/3}$ of $\bm{S}_T(t)$ exhibit a cyclic swapping property: $S_{T,\text{I}/\text{II}/\text{III}}(T) = S_{T,\text{II}/\text{III}/\text{I}}(0)$.
This property induces non-Abelian permutation of pairs of edge states, as discussed below.
\subsection{Non-Abelian permutation of three pairs of edge states}\label{Sec31}
\begin{figure}[!htp]
\includegraphics[width=1\columnwidth]{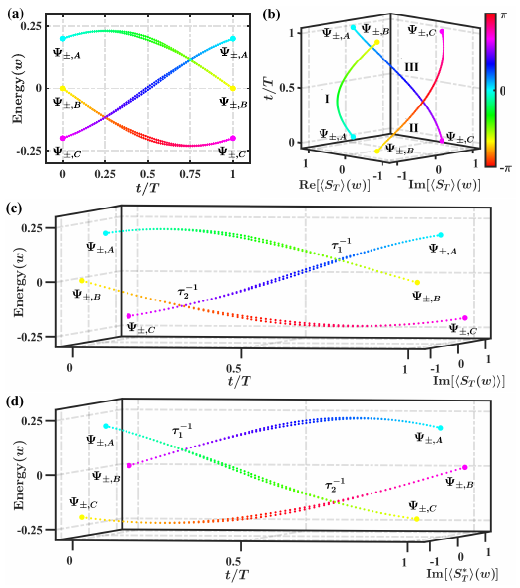}%
\caption{\label{NABraiding} 
Non-Abelian permutation of three pairs of edge states. 
(a) Energy spectrum of edge states for the triplet-coupled SSH chains. 
(b) Two branches of the expectation value $\langle\bm{S}_T\rangle$.
(c), (d) Non-Abelian braiding processes of three pairs of edge states in the ($t$, $\text{Im}[\langle\bm{S}_T\rangle]$, $E$) and ($t$, $\text{Im}[\langle\bm{S^*}_T\rangle]$, $E$) space, corresponding to the braiding operators $\tau^{-1}_{1}\tau^{-1}_{2}$ (panel (c)) and $\tau^{-1}_{2}\tau^{-1}_{1}$ (panel (d)).
In all panels, the color indicates the argument of the expectation value $\langle\bm{S}_T\rangle$ or $\langle\bm{S^*}_T\rangle$.
The parameters set as $L=14$, $w=1$, $v_{0}=0.5$, and $\eta=0.2$.}
\end{figure}

We begin with analyzing the static energy spectrum constrained by inversion symmetry $\hat{\mathcal{P}}$ and time-dependent interchain symmetry $\hat{\mathcal{S}}_{T}(t)$.
For the triple-coupled chain system, six edge states emerge in the gap, as depicted in Fig.~\ref{NABraiding}(a).
Evidently, these edge states are also common eigenstates of $\bm{P}$ and $\bm{S}_T(t)$.
Accordingly, they can be labeled as $|\Psi_{\pm,\text{I}/\text{II}/\text{III}}\rangle$ with corresponding energies $E_{\pm,\text{I}/\text{II}/\text{III}}$, where the indices are determined by the eigenvalues of the two symmetry operators.
As a result, these six edge states are grouped into three distinct branches. 
Analogous to the double-coupled case, we color-code the edge states in the spectrum of Fig.~\ref{NABraiding}(a) according to the argument of their expectation value with respect to $\bm{S}_T(t)$, denoted as $\langle\bm{S}_T\rangle = \langle\Psi|\bm{S}_T(t)|\Psi\rangle$.
As observed, in each branch the argument varies consistently with the corresponding eigenvalue $S_{T,\text{I}/\text{II}/\text{III}}$ of $\bm{S}_T(t)$.
Indeed, the real and imaginary parts of the edge states' expectation value $\langle\bm{S}_T\rangle$ form three distinct branches, which match the corresponding eigenvalues $S_{T,\text{I}/\text{II}/\text{III}}$, as shown in Fig.~\ref{NABraiding}(b).
More importantly, corresponding to the cyclic swapping property of $\bm{S}_T(t)$'s eigenvalues, the energies of these edge states exhibit an analogous behavior.
This cyclic swapping property means $S_{T,\text{I}/\text{II}/\text{III}}(T) = S_{T,\text{II}/\text{III}/\text{I}}(0)$, and the analogous behavior of energies reads $E_{\pm,\text{I}/\text{II}/\text{III}}(T)= E_{\pm,\text{II}/\text{III}/\text{I}}(0)$.
This characteristic manifests as a twisted structure of the three branches in the spectrum [Fig.~\ref{NABraiding}(a)].

What's more, during a single period, the variation of the branches (which are formed by the imaginary parts of $\langle\bm{S}_T\rangle$ and energy) is described by the non-Abelian braid operation $\tau_1^{-1}\tau_2^{-1}$, as depicted in Fig.~\ref{NABraiding}(c).
Leveraging the periodicity of both the Hamiltonian and the symmetry operator $\bm{S}_T(t)$, i.e., $\hat{H}_T(T) = \hat{H}_T(0)$ and $\bm{S}_T(T) = \bm{S}_T(0)$, we ultimately conclude that the three pairs of edge states undergo cyclic mutual transformation over one period, adhering to the relation $|\Psi_{\pm,\text{I}/\text{II}/\text{III}}(T)\rangle=|\Psi_{\pm,\text{II}/\text{III}/\text{I}}(0)\rangle$ described by the non-Abelian permutation operation $\mathcal{M}_{1}\mathcal{M}_{2}$.
To better exhibit the cyclic mutual swapping, we choose the parameters such that at both the initial and final moments of the period, the intrachain SSH segments of the system are in the fully dimerized limit ($v=0$), and the interchain couplings are decoupled ($J_{AB}=J_{AC}=J_{BC}=0$, see Eq.~\eqref{JAB}).
The edge states at these two instants are fully localized as odd-even superposition states at the two ends of individual chains, as specified below:
\begin{eqnarray}\label{PsiABC}
|\Psi_{\pm,\text{I}}(0)\rangle&=&|\Psi_{\pm,\text{III}}(T)\rangle = |\Psi_{\pm,A}\rangle, \cr\cr
|\Psi_{\pm,\text{II}}(0)\rangle&=&|\Psi_{\pm,\text{I}}(T)\rangle = |\Psi_{\pm,B}\rangle,\cr\cr
|\Psi_{\pm,\text{III}}(0)\rangle&=&|\Psi_{\pm,\text{II}}(T)\rangle = |\Psi_{\pm,C}\rangle,\cr\cr
|\Psi_{\pm,A/B/C}\rangle&=&\frac{|1_{A/B/C}\rangle \pm |L_{A/B/C}\rangle}{\sqrt{2}}.
\end{eqnarray}
The cyclic mutual swapping of the edge states can be straightforwardly confirmed by the exchange of their positions among Chains $A$, $B$, and $C$.

The quintessential effect of non-Abelian permutation manifests as the operation-sequence-dependent swapping of edge states.
To this end, we perform a time reversal operation on the Hamiltonian $\hat{H}_T(t)$ and the symmetry operator $\hat{\mathcal{S}}_T(t)$, that is
\begin{eqnarray}\label{TimeReversal}
\hat{H}_T(t)\rightarrow\hat{H}^*_T(-t),~~
\hat{\mathcal{S}}_T(t)\rightarrow\hat{\mathcal{S}}^*_T(-t).
\end{eqnarray}
As a result, during a single period, the variation of the branches exhibits a distinct pattern.
Specifically, for branches formed by the imaginary parts of $\langle\bm{S}^*_T\rangle$ and energy, this variation is described by the non-Abelian braid operation $\tau_2^{-1}\tau_1^{-1}$, as depicted in Fig.~\ref{NABraiding}(d).
The cyclic mutual swapping of the edge states is distinctly different from that described by $\mathcal{M}_{1}\mathcal{M}_{2}$ (see Eq.~\eqref{PsiABC}) and is given by
\begin{eqnarray}\label{PsiACB}
|\Psi_{\pm,\text{I}}(0)\rangle&=&|\Psi_{\pm,\text{II}}(T)\rangle = |\Psi_{\pm,A}\rangle, \cr\cr
|\Psi_{\pm,\text{II}}(0)\rangle&=&|\Psi_{\pm,\text{III}}(T)\rangle = |\Psi_{\pm,B}\rangle,\cr\cr
|\Psi_{\pm,\text{III}}(0)\rangle&=&|\Psi_{\pm,\text{I}}(T)\rangle = |\Psi_{\pm,C}\rangle.
\end{eqnarray}
Obviously, after traversing a period, starting with the same initial state, the final state governed by $\mathcal{M}_{2}\mathcal{M}_{1}$ differs from that governed by $\mathcal{M}_{1}\mathcal{M}_{2}$.
This is precisely the manifestation of non-Abelian permutation. 
In this subsection, we have discussed symmetry-protected non-Abelian permutation; below, we will demonstrate non-Abelian topological transport based on this non-Abelian permutation.
\subsection{Non-Abelian topological transport}\label{Sec32}
As discussed in the double-coupled system, particles initially injected into one end of a chain can be transferred to the two ends of the other chain at a tunable ratio ranging from $0$ to $1$, depending on the relative dynamical phase.
In this subsection, we will illustrate more intricate non-Abelian topological transport among three chains.
Without loss of generality, we initiate the adiabatic evolution of the edge states under symmetry protection by considering an initial state at the left end of Chain B, which can be expressed as a superposition of edge states in Branch II:
\begin{eqnarray}\label{Evot}
|\varPhi(t)\rangle &=& \mathcal{T}e^{-i\int_{0}^{t}\hat{H}_T(\tau)d\tau}|\varPhi(0)\rangle,  \cr\cr
|\varPhi(0)\rangle &=& |1_B\rangle = \frac{|\Psi_{+,\text{II}}(0)\rangle + |\Psi_{-,\text{II}}(0)\rangle}{\sqrt{2}}.
\end{eqnarray}
Analogous to the double-coupled case, the symmetry-protected adiabatic evolution exhibits two key features: first, the time-dependent interchain symmetry $\hat{\mathcal{S}}_{T}(t)$ ensures that $|\varPhi(t)\rangle$ stays within the initial branch during evolution at sufficiently slow velocity; second, inversion symmetry $\hat{\mathcal{P}}$ leads to the independent evolution of components with different parities in the initial state.
Thereby, following a period of adiabatic evolution, the final state can be expressed up to a global phase factor as:
\begin{eqnarray}\label{Evoend}
|\varPhi(T)\rangle &=& \frac{e^{-i\frac{\theta_d}{2}}|\Psi_{+,\text{II}}(T)\rangle + e^{i\frac{\theta_d}{2}}|\Psi_{-,\text{II}}(T)\rangle}{\sqrt{2}} \cr\cr
&=& \cos\frac{\theta_d}{2}|1_C\rangle -i\sin\frac{\theta_d}{2}|L_C\rangle,
\end{eqnarray}
where $\theta_d = \int_{0}^{T} \left[ E_{+,\text{II}}(t) - E_{-,\text{II}}(t) \right] dt$ is the accumulated dynamical phase difference.
The second equality leverages the relationships presented in Eq.~\eqref{PsiABC}, which are associated with the non-Abelian permutation process $\mathcal{M}_{1}\mathcal{M}_{2}$.
Notably, this result shows that particles injected initially at the left end of Chain $B$ can be transferred to the two ends of Chain $A$ with a continuously tunable ratio (0 to 1), governed by the relative dynamical phase $\theta_d$.
Similarly, for the initial state $|L_{B}\rangle$, we observe the symmetric behavior.
Furthermore, when the initial state is placed at one end of the other chains, the results exhibit cyclic mutual transport, both of which are summarized as follows
\begin{eqnarray}\label{BS2}
|1_{A/B/C}\rangle &\rightarrow \cos\frac{\theta_{d}}{2}|1_{B/C/A}\rangle -i\sin\frac{\theta_{d}}{2}|L_{B/C/A}\rangle, \cr\cr
|L_{A/B/C}\rangle &\rightarrow -i\sin\frac{\theta_{d}}{2}|1_{B/C/A}\rangle + \cos\frac{\theta_{d}}{2}|L_{B/C/A}\rangle.
\end{eqnarray}
The neglected global phase differs for each pair of edge states, yet the relative dynamical phases $\theta_{d}$ for each pair during one period are the same:
$\int_{0}^{T} \left[ E_{+,\text{I}}(t) - E_{-,\text{I}}(t) \right] dt=\int_{0}^{T} \left[ E_{+,\text{II}}(t) - E_{-,\text{II}}(t) \right] dt=\int_{0}^{T} \left[ E_{+,\text{III}}(t) - E_{-,\text{III}}(t) \right] dt$.
When considering non-Abelian transport governed by the permutation operator  $\mathcal{M}_{2}\mathcal{M}_{1}$, the cyclic mutual transport takes a distinct form as follows:
\begin{eqnarray}\label{BS3}
|1_{A/B/C}\rangle&\rightarrow \cos\frac{\theta_{d}}{2}|1_{C/A/B}\rangle -i\sin\frac{\theta_{d}}{2}|L_{C/A/B}\rangle, \cr\cr
|L_{A/B/C}\rangle&\rightarrow -i\sin\frac{\theta_{d}}{2}|1_{C/A/B}\rangle + \cos\frac{\theta_{d}}{2}|L_{C/A/B}\rangle.
\end{eqnarray}

\begin{figure}[!htp]
\includegraphics[width=1\columnwidth]{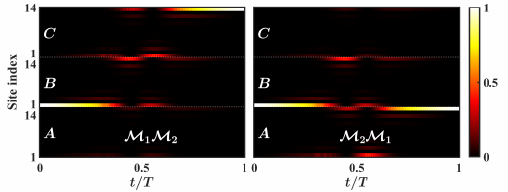}
\caption{\label{Transport}
Non-Abelian topological transport. Density distribution $\langle\hat{n}_{j}\rangle$ as a function of the evolution time $t$ for the permutation operator $\mathcal{M}_{1}\mathcal{M}_{2}$ (left) and  $\mathcal{M}_{2}\mathcal{M}_{1}$ (right). The initial state is $|1_{B}\rangle$, with parameters set as $L=14$, $w=1$, $v_{0}=0.5$, $\eta=0.2$, and $T=2663$ (corresponding to $\theta_{d}=3\pi$).}
\end{figure}

We present the density distribution $\langle\hat{n}_{j}\rangle$ as a function of the evolution time $t$.
Here, we select the initial state as $|1_{B}\rangle$, with the system parameters set to period $T=2663$ and chain length $L=14$ (corresponding to $\theta_{d}=3\pi$).
Specifically, as shown in Fig.~\ref{Transport}, the final density distribution localizes at the right end of Chain $C$ for the permutation operator $\mathcal{M}_{1}\mathcal{M}_{2}$, while it localizes at the right end of Chain $A$ for $\mathcal{M}_{2}\mathcal{M}_{1}$.
It intuitively demonstrates the non-Abelian nature of topological transport, characterized by an operation-sequence-dependent outcome.

Actually, there has been extensive and comprehensive research on state permutations, band braiding, and non-Abelian state transfer in non-Hermitian systems \cite{HuPRL2021,HuPRR2022,GuoPRL2023,WangNanotechnology2025,TangNSR2022,BaoPRB2024}.
However, in these works, the demonstration of non-Abelian properties is often reliant on quasi-static processes, which imposes limitations on their practical applications.
In this work, we overcome this limitation by implementing non-Abelian permutation protected by a time-dependent non-Hermitian symmetry, rather than by the  Hamiltonian itself.
Looking forward, we can build on these works to extend our results to systems with more state permutations or braiding branches, explore their relevance to practical quantum computation, and develop a more comprehensive theoretical framework.
\subsection{Non-Abelian Hong-Ou-Mandel interference}\label{Sec33}
As discussed in Sec.~\ref{Sec22} and \ref{Sec32}, placing the initial state at one end of a chain results in its transfer to the two ends of other chains, with a continuously tunable ratio (from 0 to 1) determined by the relative dynamical phase difference accumulated over an adiabatic evolution period.
Controlling this tunable ratio offers numerous promising applications in quantum science and technology, including the non-Abelian topological transport (operating at an extreme 0:100 ratio).
In this subsection, we focus on non-Abelian HOM interference with a 50:50 ratio.

Consider two identical bosons injected into the two ends of Chain $B$, with their initial two-boson state given by $|1_{B},L_{B}\rangle$.
When subjected to an adiabatic evolution period governed by the permutation operator $\mathcal{M}_{1}\mathcal{M}_{2}$, these bosons transform according to Eq.~\ref{BS2}.
Specifically, the initial state $|1_{B},L_{B}\rangle$ evolves into 
$-i\sin\theta_{d}(|1_{C},1_{C}\rangle+|L_{C},L_{C}\rangle)/\sqrt{2}+\cos\theta_{d} |1_{C},L_{C}\rangle$.
We choose $\theta_{d} = \frac{2n+1}{2}\pi$ ($n \in \mathbb{Z}$), which corresponds to a 50:50 ratio; under this condition, the antibunching term vanishes.
In this case, the state simplifies to the two-particle NOON state
\begin{eqnarray}\label{PhiNOONC} 
|\Psi_{{\rm NOON},C}\rangle = \frac{1}{\sqrt{2}}\left(|1_{C},1_{C}\rangle + |L_{C},L_{C}\rangle\right),
\end{eqnarray}
neglecting a global phase factor.
As a maximally entangled state between the two ends of Chain $C$, this spatial NOON state fundamentally arises from the well-known HOM interference.
Notably, the HOM interference exhibits non-Abelian nature; when subjected to adiabatic evolution governed by the permutation operator $\mathcal{M}_{2}\mathcal{M}_{1}$, the bosons transform according to Eq.~\ref{BS3}, resulting in the final NOON state belonging to Chain $A$
\begin{eqnarray}\label{PhiNOONA} 
|\Psi_{{\rm NOON},A}\rangle = \frac{1}{\sqrt{2}}\left(|1_{A},1_{A}\rangle + |L_{A},L_{A}\rangle\right).
\end{eqnarray}

\begin{figure}[!htp]
\includegraphics[width=1\columnwidth]{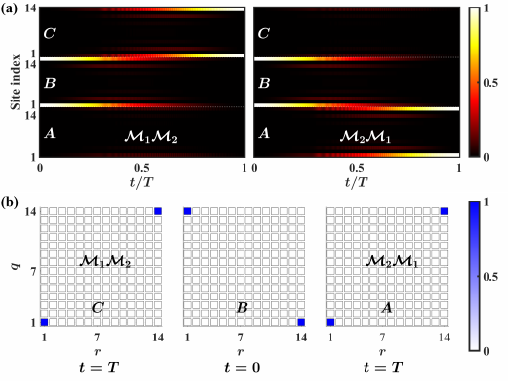}
\caption{\label{HOM}Non-Abelian HOM interference. (a) Density distribution $\langle\hat{n}_{j}\rangle$ as a function of evolution time $t$ for the permutation operators $\mathcal{M}_{1}\mathcal{M}_{2}$ (left) and $\mathcal{M}_{2}\mathcal{M}_{1}$ (right). (b) Two-particle correlation $\Gamma_{q,r}$ at initial and final times for the same permutation operators [$\mathcal{M}_{1}\mathcal{M}_{2}$ (left) and $\mathcal{M}_{2}\mathcal{M}_{1}$ (right)]. The initial state is $|1_{B},L_{B}\rangle$, with parameters set as $L=14$, $w=1$, $v_{0}=0.5$, $\eta=0.2$, and $T=2220$ (corresponding to $\theta_{d}=2.5\pi$).}
\end{figure}

To intuitively illustrate the non-Abelian nature of HOM interference (i.e., operation-sequence-dependent outcomes), we present the density distribution $\langle\hat{n}_{j}\rangle$ over evolution time $t$ during adiabatic evolution in Fig.~\ref{HOM}(a).
We initialize the system with the product state $|1_{B},L_{B}\rangle$, fixing system parameters to period $T=2220$ and chain length $L=14$ (corresponding to $\theta_{d}=2.5\pi$).
As shown in Fig.~\ref{HOM}(a), this leads to a clear contrast: the final density distribution is uniformly distributed at Chain $C$'s two ends for $\mathcal{M}_{1}\mathcal{M}_{2}$ (left), while it localizes at Chain $A$'s two ends for $\mathcal{M}_{2}\mathcal{M}_{1}$ (right). 
This directly reflects the non-Abelian nature.

To characterize the correlation properties of the entanglement in the generated two-particle NOON state, we compute the two-particle correlation $\Gamma_{q,r} = \langle\hat{b}_{q}^{\dag}\hat{b}_{r}^{\dag}\hat{b}_{r}\hat{b}_{q}\rangle$ and plot it in Fig.~\ref{HOM}(b).
At the initial time $t=0$, the correlation function shows an uncorrelated antibunching pattern localized on Chain $B$.
Following one adiabatic period, a distinct transition occurs: for the permutation operator $\mathcal{M}_{1}\mathcal{M}_{2}$, the pattern evolves into a correlated bunching distribution at Chain $C$'s two ends (left); for $\mathcal{M}_{2}\mathcal{M}_{1}$, it localizes at Chain $A$'s two ends (right).
Notably, the fidelity between the final state and the target state ($|\Psi_{{\rm NOON},C}\rangle$ or $|\Psi_{{\rm NOON},A}\rangle$) reaches 0.995 in both cases.
Thus, we successfully realize the spatially entangled NOON state via non-Abelian HOM interference.

\section{SUMMARY AND DISCUSSION}\label{Sec4}
In this article, we have systematically investigated the permutation of topological edge state pairs protected by dual symmetries, namely time-independent inversion symmetry and time-dependent interchain symmetry, based on coupled SSH chains.
Specifically, we observed the twisted energy spectrum and braiding structure of these edge state pairs, which originate from the swapping property of the eigenvalues of the time-dependent interchain symmetry operator.
Furthermore, we achieved symmetry-protected adiabatic evolution and realized high-fidelity, ratio-tunable topological transport from one end of a chain to the two ends of another chain.
Through non-Abelian topological transport and HOM interference of edge states, we demonstrate the key signatures of non-Abelian permutation: specifically, the swapping of edge state pairs and the order dependence of permutation outcomes.
This symmetry-protected non-Abelian transport and interference can be directly generalized to scenarios with more than three pairs of edge states by designing appropriate symmetry operators and Hamiltonians \cite{LongPRL2024}. 
Additionally, considering internal degrees of freedom such as spin may eliminate the need for next-nearest coupling required for the permutation of multiple pairs of edge states.
Moreover, we point out that the symmetry-protected non-Abelian permutation can be extended to bulk states.

Lastly, we briefly discuss the experimental feasibility of our protocol.
Photonic waveguide arrays represent a promising platform for exploring topological features and non-Abelian physics.
Indeed, a range of key phenomena, including adiabatic pumping of topological edge states \cite{YEKrausPRL2012}, topological protection of entanglement \cite{MCRechtsmanOptica2016,MWangNanophotonics2019}, HOM interference of topological states of light \cite{JLTambascoSA2018}, non-Abelian Thouless pumping \cite{SunNP2022}, and non-Abelian braiding of photonic modes \cite{ZhangNP2022}have been observed using this system.
Notably, the adiabatically modulated Aubry-Andr\'{e}-Harper model can be realized by slowly varying the spacing between waveguides along the propagation axis \cite{YEKrausPRL2012,JLTambascoSA2018}.
Thus, adiabatically modulated photonic waveguide arrays serve as a viable platform for testing our protocol.

\acknowledgements{The authors thank L. Li for helpful discussions. This work is supported by the National Natural Science Foundation of China under Grant No. 12104103.}

\section*{Data availability}
The data that support the findings of this article are not publicly available. The data are available from the authors upon reasonable request.

\appendix
\renewcommand\theequation{\Alph{section}\arabic{equation}}

\section{Time dependent symmetry and adiabatic
evolution velocity}\label{appA}
In this section, we investigate adiabatic evolution under the protection of two symmetries: time-independent inversion symmetry and time-dependent interchain symmetry.
We focus on the double-coupled case without loss of generality, with the results generalizable to the triplet-coupled system.
We initialize the system with an even-parity edge state in Branch I, whose initial state is given by
\begin{eqnarray}\label{A1}
|\varphi(0)\rangle = |\psi_{+,\text{I}}(0)\rangle =  \frac{|1_{A}\rangle + |L_{A}\rangle}{\sqrt{2}}.
\end{eqnarray}
Consistent with Sec.~\ref{Sec2}, we adopt the same parameter variations; additionally, we tune the evolution velocity by adjusting the period duration $T$.
 
\begin{figure}[!htp]
\includegraphics[width=1\columnwidth]{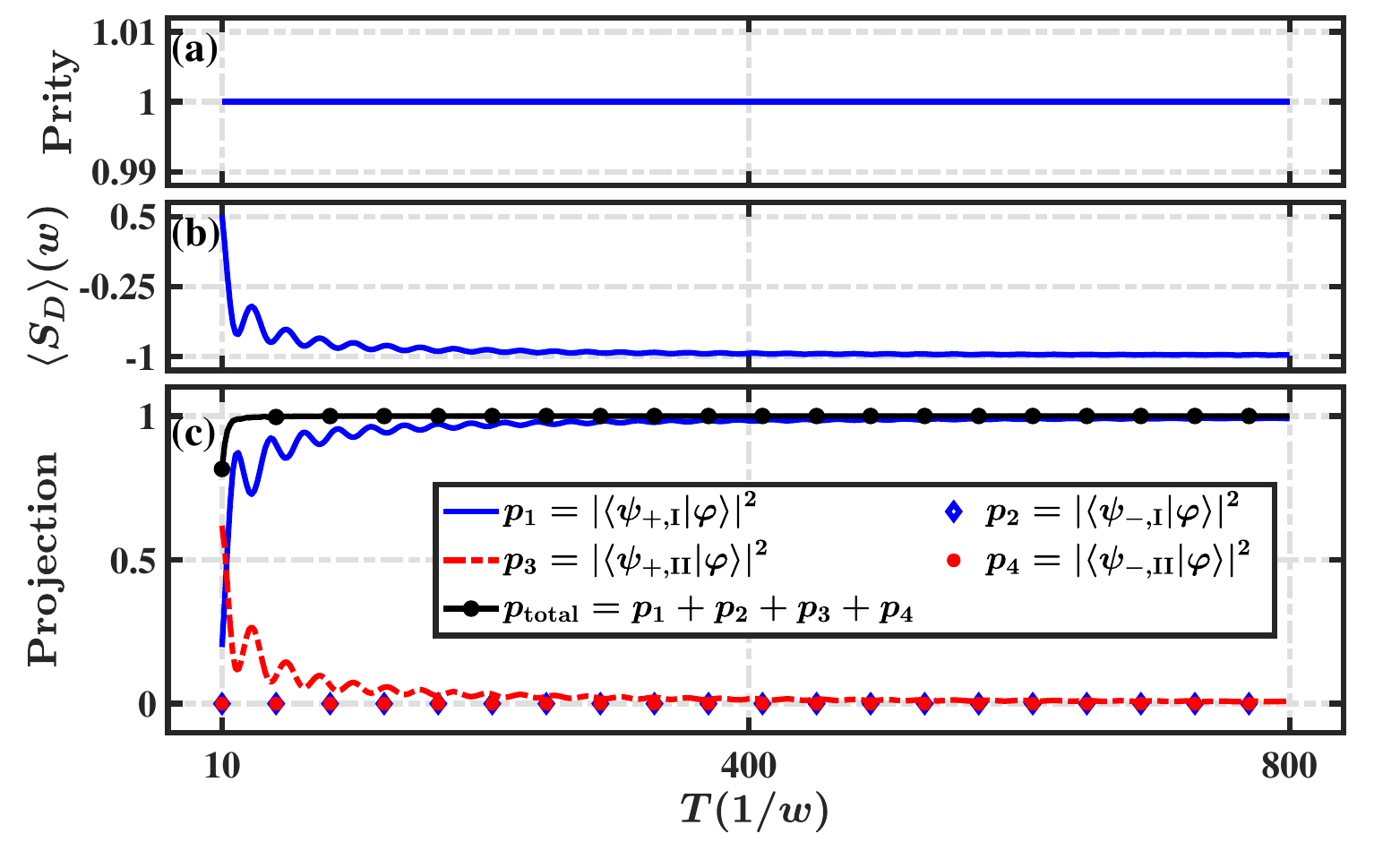}%
\caption{\label{Adiabatic}(a) Parity; (b) Expectation value $\langle\bm{S}_D\rangle$; (c) Projection probabilities versus the period duration $T$. All quantities are computed based on the final state after one evolution period. The initial state is $|\psi_{+,\text{I}}(0)\rangle$, with parameters set as $L=14$, $w=1$, $v_{0}=0.5$, and $\kappa=0.2$.}
\end{figure}

Time-independent inversion symmetry $\hat{\mathcal{P}}$ serves as the fundamental guarantee for parity conservation in the system.
As shown in Fig.~\ref{Adiabatic}(a), the final state retains a parity of $p=1$ after one evolution period, regardless of the period duration $T$.
This observation directly confirms the parity conservation enforced by $\hat{\mathcal{P}}$.
In contrast, the expectation value of $\bm{S}_D(t)$ after one evolution period ($\langle\bm{S}_D\rangle = \langle\varphi(T)|\bm{S}_D(T)|\varphi(T)\rangle$) shows a clear dependence on $T$.
Specifically, as $T$ increases, $\langle\bm{S}_D\rangle$ asymptotically converges to the eigenvalue $S_{D,\text{I}}(T) = -1$ [see Fig.~\ref{Adiabatic}(b)].
To quantify the distribution of the final state $|\varphi(T)\rangle$ across distinct edge states, we compute the projection probabilities onto these states, defined specifically as:
\begin{eqnarray}\label{A2}
&p_1= |\langle\psi_{+,\text{I}}(T)|\varphi(T)\rangle|^2,~~ 
p_2 = |\langle\psi_{-,\text{I}}(T)|\varphi(T)\rangle|^2, \cr\cr
&p_3= |\langle\psi_{+,\text{II}}(T)|\varphi(T)\rangle|^2,~~ 
p_4 = |\langle\psi_{-,\text{II}}(T)|\varphi(T)\rangle|^2,\cr
\end{eqnarray}
with the total projection probability $p_{\rm total} = \sum_{i=1}^4 p_i$.
As shown in Fig.~\ref{Adiabatic}(c), due to parity conservation, the projection probabilities $p_2$ (blue diamonds) and $p_4$ (red dots) remain strictly zero for all values of $T$.
The total projection $p_{\rm total}$ (black dot-marked line) quickly approaches unity as $T$ increases, indicating near-complete projection onto the considered edge states.
Notably, however, $p_1$ (blue line) requires a sufficiently long $T$ to converge to 1.
In conclusion, in stark contrast to time-independent inversion symmetry (which imposes strict, $T$-independent constraints), adiabatic evolution protected by time-dependent interchain symmetry demands a sufficiently long evolution time, though it is not of infinite duration.


\begin{thebibliography}{99}
\bibitem{MZHasanRMP2010}
M. Z. Hasan and C. L. Kane, Colloquium: Topological insulators, Rev. Mod. Phys. \textbf{82}, 3045 (2010).
\bibitem{XLQiRMP2011}
X.-L. Qi and S.-C. Zhang, Topological insulators and superconductors, Rev. Mod. Phys. \textbf{83}, 1057 (2011).
\bibitem{CNayakRMP2008}
C. Nayak, S. H. Simon, A. Stern, M. Freedman, and S. Das Sarma, Non-Abelian anyons and topological quantum computation, Rev. Mod. Phys. \textbf{80}, 1083 (2008).
\bibitem{RMLutchynPRL2010}
R. M. Lutchyn, J. D. Sau, and S. Das Sarma, Majorana Fermions and a Topological Phase Transition in Semiconductor-Superconductor Heterostructures, Phys. Rev. Lett. \textbf{105}, 077001 (2010).

\bibitem{KlitzingPRL1980}
K. v. Klitzing, G. Dorda, andM. Pepper, NewMethod for High-Accuracy Determination of the Fine-Structure Constant Based on Quantized Hall Resistance, Phys. Rev. Lett. \textbf{45}, 494 (1980).
\bibitem{ThoulessPRL1982}
D. J. Thouless, M. Kohmoto, M. P. Nightingale, and M. den Nijs, Quantized Hall Conductance in a Two-Dimensional Periodic Potential, Phys. Rev. Lett. \textbf{49}, 405 (1982).
\bibitem{KanePRL2005}
C. L. Kane and E. J. Mele, $Z_2$ Topological Order and the Quantum Spin Hall Effect, Phys. Rev. Lett. \textbf{95}, 146802 (2005).
\bibitem{KaneGraphenePRL2005}
C. L. Kane and E. J. Mele, Quantum Spin Hall Effect in Graphene, Phys. Rev. Lett. \textbf{95}, 226801 (2005).
\bibitem{BernevigPRL2006}
B. A. Bernevig and S.-C. Zhang, Quantum Spin Hall Effect, Phys. Rev. Lett. \textbf{96}, 106802 (2006).
\bibitem{MourikScience2012}
V. Mourik, K. Zuo, S. M. Frolov, S. R. Plissard, E. P. A. M. Bakkers, and L. P. Kouwenhoven, Signatures of Majorana fermions in hybrid superconductor-semiconductor nanowire devices, Science \textbf{336}, 1003 (2012).
\bibitem{DasNP2012}
A. Das, Y. Ronen, Y. Most, Y. Oreg, M. Heiblum, and H. Shtrikman, Zero-bias peaks and splitting in an Al-InAs nanowire topological superconductor as a signature of Majorana fermions, Nat. Phys. \textbf{8}, 887 (2012).
\bibitem{XuScience2015}
S.-Y. Xu, I. Belopolski, N. Alidoust, M. Neupane, G. Bian, C. Zhang, R. Sankar, G. Chang, Z. Yuan, C.-C.g Lee, S.-M. Huang, H. Zheng, J. Ma, D. S. Sanchez, B. Wang, A. Bansil, F. Chou, P. P. Shibayev, H. Lin, S. Jia, and M. Z. Hasan, Discovery of a Weyl fermion semimetal and topological Fermi arcs, Science \textbf{349}, 613 (2015).
\bibitem{SongPRX2019}
Z. Song and X. Dai, Hear the sound of Weyl fermions, Phys. Rev. X \textbf{9}, 021053 (2019).

\bibitem{YHatsugaiPRL1993}
Y. Hatsugai, Chern number and edge states in the integer quantum Hall effect, Phys. Rev. Lett. \textbf{71}, 3697 (1993).
\bibitem{EssinPRB2011}
A. M. Essin and V. Gurarie, Bulk-boundary correspondence of topological insulators from their respective green’s functions, Phys. Rev. B \textbf{84}, 125132 (2011).

\bibitem{RyuNJP2010}
S. Ryu, A. P. Schnyder, A. Furusaki, and A. W. W. Ludwig, Topological insulators and superconductors: tenfold way and dimensional hierarchy, New J. Phys. \textbf{12}, 065010 (2010).
\bibitem{CKChiuRMP2016}
C.-K. Chiu, J. C. Y. Teo, A. P. Schnyder, and S. Ryu, Classification of topological quantum matter with symmetries, Rev. Mod. Phys. \textbf{88}, 035005 (2016).

\bibitem{ZWangNature2009}
Z. Wang, Y. Chong, J. D. Joannopoulos, and M. Soljačić, Observation of unidirectional backscattering-immune topological electromagnetic states. Nature \textbf{461}, 772 (2009).
\bibitem{JSeoNature2010}
J. Seo, P. Roushan, H. Beidenkopf, Y. S. Hor, R. J. Cava, and A. Yazdani, Transmission of topological surface states through surface barriers, Nature \textbf{466}, 343 (2010).


\bibitem{YEKrausPRL2012}
Y. E. Kraus, Y. Lahini, Z. Ringel, M. Verbin, and O. Zilberberg, Topological States and Adiabatic Pumping in Quasicrystals, Phys. Rev. Lett. \textbf{109}, 
\bibitem{NLangQI2017}
N. Lang and H. P. Büchler, Topological networks for quantum communication between distant qubits, npj Quantum Inf. \textbf{3}, 47 (2017).
\bibitem{CDlaskaQST2017}
C. Dlaska, B. Vermersch, and P. Zoller, Robust quantum state transfer via topologically protected edge channels in dipolar arrays, Quantum Sci. Technol. \textbf{2}, 015001 (2017).
\bibitem{FMeiPRA2018}
F. Mei, G. Chen, L. Tian, S. L. Zhu, and S. Jia, Robust quantum state transfer via topological edge states in superconducting qubit chains, Phys. Rev. A \textbf{98}, 012331 (2018).
\bibitem{SLonghiPRB2019}
S. Longhi, Topological pumping of edge states via adiabatic passage, Phys. Rev. B \textbf{99}, 155150 (2019).
\bibitem{NEPalaiodimopoulosPRA2021}
N. E. Palaiodimopoulos, I. Brouzos, F. K. Diakonos, and G.Theocharis, Fast and robust quantum state transfer via a topological chain, Phys. Rev. A \textbf{103}, 052409 (2021).
\bibitem{LHuangPRA2022}
L. Huang, Z. Tan, H. Zhong, and B. Zhu, Fast and robust quantum state transfer assisted by zero-energy interface states in a splicing Su-Schrieffer-Heeger chain, Phys. Rev. A \textbf{106}, 022419 (2022).
\bibitem{CWangPRA2022}
C. Wang, L. Li, J. Gong, and Y.-X. Liu, Arbitrary entangled state transfer via a topological qubit chain, Phys. Rev. A \textbf{106}, 052411 (2022).


\bibitem{PBorossPRB2019}
P. Boross, J. K. Asbóth, G. Széchenyi, L. Oroszlány, and A. Pályi, Poor man’s topological quantum gate based on the Su-Schrieffer-Heeger model, Phys. Rev. B \textbf{100}, 045414 (2019).
\bibitem{MNarozniakPRB2021}
M. Naro\.{z}niak, M. C. Dartiailh, J. P. Dowling, J. Shabani, and T. Byrnes, Quantum gates for Majoranas zero modes in topological superconductors in one-dimensional geometry, Phys. Rev. B
\textbf{103}, 205429 (2021).


\bibitem{RHammerPRB2013}
R. Hammer and W. P\"{o}tz, Dynamics of domain-wall Dirac fermions on a topological insulator: A chiral fermion beam splitter, Phys. Rev. B \textbf{88}, 235119 (2013).
\bibitem{XSWangPRB2017}
X. S. Wang, Y. Su, and X. R. Wang, Topologically protected unidirectional edge spin waves and beam splitter, Phys. Rev. B \textbf{95}, 014435 (2017).
\bibitem{LQiPRB2021}
L. Qi, Y. Xing, X. D. Zhao, S. Liu, S. Zhang, S. Hu, and H. F. Wang, Topological beam splitter via defect-induced edge channel in the Rice-Mele model, Phys. Rev. B \textbf{103}, 085129 (2021).
\bibitem{LQiPRA2023}
L. Qi, N. Han, S. Hu, and A.-L. He, Engineering the unidirectional topological excitation transmission and topological diode in the Rice-Mele model, Phys. Rev. A \textbf{108}, 032402 (2023).

\bibitem{AEkertRMP1996}
A. Ekert and R. Jozsa, Quantum computation and Shor’s factoring algorithm, Rev. Mod. Phys. \textbf{68}, 733 (1996).
\bibitem{JWPanRMP2012}
J.-W. Pan, Z.-B. Chen, C.-Y. Lu, H. Weinfurter, A. Zeilinger, and M. Zukowski, Multiphoton entanglement and interferometry, Rev. Mod. Phys. \textbf{84}, 777 (2012).
\bibitem{LPezzeRMP2018}
L. Pezz\`{e}, A. Smerzi, M. K. Oberthaler, R. Schmied, and P. Treutlein, Quantum metrology with nonclassical states of atomic ensembles, Rev. Mod. Phys. \textbf{90}, 035005 (2018).

\bibitem{MCRechtsmanOptica2016}
M. C. Rechtsman, Y. Lumer, Y. Plotnik, A. Perez-Leija, A. Szameit, and M. Segev, Topological protection of photonic path entanglement, Optica \textbf{3}, 925 (2016).
\bibitem{MWangNanophotonics2019}
M. Wang, C. Doyle, B. Bell, M. J. Collins, E. Magi, B. J. Eggleton, M. Segev, and A. Blanco-Redondo, Topologically protected entangled photonic states, Nanophotonics \textbf{8}, 1327 (2019).
\bibitem{KMonkmanPRR2020}
K. Monkman and J. Sirker, Operational entanglement of symmetry-protected topological edge states, Phys. Rev. Res. \textbf{2}, 043191 (2020).
\bibitem{JXHanPRA2021}
J.-X. Han, J.-L. Wu, Y. Wang, Y. Xia, Y.-Y. Jiang, and J. Song, Large-scale Greenberger-Horne-Zeilinger states through a topologically protected zero-energy mode in a superconducting qutrit-resonator chain, Phys. Rev. A \textbf{103}, 032402 (2021).


\bibitem{CKHongPRL1987}
C. K. Hong, Z. Y. Ou, and L. Mandel, Measurement of subpicosecond time intervals between two photons by interference, Phys. Rev. Lett. \textbf{59}, 2044 (1987).
\bibitem{JLTambascoSA2018}
J. L. Tambasco, G. Corrielli, R. J. Chapman, A. Crespi, O. Zilberberg, R. Osellame, and A. Peruzzo, Quantum interference of topological states of light, Sci. Adv. \textbf{4}, eaat3187 (2018).

\bibitem{SHuPRA2020}
S. Hu, Y. Ke, and C. Lee, Topological quantum transport and spatial entanglement distribution via a disordered bulk channel, Phys. Rev. A \textbf{101}, 052323 (2020).
\bibitem{SHuPRA2024}
S. Hu, M. Hu, S. Li, Z. Zhong, and Z. Lei, Hong-Ou-Mandel interference in a temporal-average-inversion-symmetric chain, Phys. Rev. A \textbf{110}, 032438 (2024).
\bibitem{SHuPRA2025}
S. Hu, S. Li, M. Hu, and Z. Lei, Symmetry-protected Landau-Zener-Stückelberg-Majorana interference and nonadiabatic topological transport of edge states, Phys. Rev. A \textbf{111}, 052414 (2025).

\bibitem{KitaevAP2003}
A. Kitaev, Fault-tolerant quantum computation by anyons, Ann. Phys. \textbf{303}, 2 (2003).
\bibitem{AliceaNP2011}
J. Alicea, Y. Oreg, G. Refael, F. von Oppen, and M. P. A. Fisher, Non-Abelian statistics and topological quantum information processing in 1D wire networks, Nat. Phys. \textbf{7}, 412 (2011).
\bibitem{Sarmanpj2011}
S. D. Sarma, M. Freedman, and C. Nayak, Majorana zero modes and topological quantum computation, npj Quantum Inf. \textbf{1}, 15001 (2015).
\bibitem{WuScience2019}
Q. Wu, A. A. Soluyanov, and T. Bzdušek, Non-Abelian band topology in noninteracting metals, Science \textbf{365}, 1273–1277 (2019).
\bibitem{BouhonNP2020}
A. Bouhon, Q. Wu, R.-J. Slager, H. Weng, O. V. Yazyev, and T. Bzdušek, Non-Abelian reciprocal braiding of Weyl points and its manifestation in ZrTe, Nat. Phys. \textbf{16}, 1137–1143 (2020).
\bibitem{BouhonPRB2020}
A. Bouhon, T. Bzdušek, and R.-J. Slager, Geometric approach to fragile topology beyond symmetry indicators, Phys. Rev. B \textbf{102}, 115135 (2020).
\bibitem{BouhonPRB2021}
A. Bouhon, G. F. Lange, and R.-J. Slager, Topological correspondence between magnetic space group representations and subdimensions, Phys. Rev. B \textbf{103}, 245127 (2021).
\bibitem{GuoNature2021}
Q. Guo, T. Jiang, R.-Y. Zhang, L. Zhang, Z.-Q. Zhang, B. Yang, S. Zhang, and C. T. Chan, Experimental observation of non-Abelian topological charges and edge states, Nature \textbf{594}, 195–200 (2021).
\bibitem{BroscoPRA2021}
V. Brosco, L. Pilozzi, R. Fazio, and C. Conti, Non-Abelian Thouless pumping in a photonic lattice, Phys. Rev. A \textbf{103}, 063518 (2021).
\bibitem{SunNP2022}
Y.-K. Sun, X.-L. Zhang, F. Yu, Z.-N. Tian, Q.-D. Chen and H.-B. Sun, Non-Abelian Thouless pumping in photonic waveguides, Nat. Phys. \textbf{18}, 1080–1085 (2022).
\bibitem{PengNC2022}
B. Peng, A. Bouhon, B. Monserrat, and R.-J. Slager, Phonons as a platform for non-Abelian braiding and its manifestation in layered silicates, Nat. Commun. \textbf{13}, 423 (2022). 
\bibitem{LiNC2023}
T. Li and H. Hu, Floquet non-Abelian topological insulator and multifold bulk-edge correspondence, Nat. Commun. \textbf{14}, 6418 (2023). 
\bibitem{PengNC2024}
R.-J. Slager, A. Bouhon, and F. N. Ünal, Non-Abelian Floquet braiding and anomalous Dirac string phase in periodically driven systems, Nat. Commun. \textbf{15}, 1144 (2024).
\bibitem{SunPRL2024}
X.-C. Sun, J.-B. Wang, C. He, and Y.-F. Chen, Non-Abelian Topological Phases and Their Quotient Relations in Acoustic Systems, Phys. Rev. Lett. \textbf{132}, 216602 (2024).
\bibitem{BreachPRL2024}
O. Breach, R.-J. Slager, and F. N. Ünal, Interferometry of Non-Abelian Band Singularities and Euler Class Topology, Phys. Rev. Lett. \textbf{133}, 093404 (2024).
\bibitem{ChenPRR2025}
L. Chen, M. Fuertes, and B. Deng, Topological computation by non-Abelian braiding in classical metamaterials, Phys. Rev. Res. \textbf{7}, 023143 (2025).



\bibitem{YXZhaoNC2022}
Z. Y. Chen, Shengyuan A. Yang, and Y. X. Zhao, Brillouin Klein bottle from artificial gauge fields, Nat. Comm. \textbf{13}, 2215 (2022).
\bibitem{XuePRL2022}
H. Xue, Z. Wang, Y.-X. Huang, Z. Cheng, L. Yu, Y. X. Foo, Y. X. Zhao, S. A. Yang, and B. Zhang, Projectively enriched symmetry and topology in acoustic crystals, Phys. Rev. Lett. \textbf{128}, 116802 (2022).
\bibitem{LiPRL2022}
T. Li, J. Du, Q. Zhang, Y. Li, X. Fan, F. Zhang, and C. Qiu, Acoustic möbius insulators from projective symmetry, Phys. Rev. Lett. \textbf{128}, 116803 (2022).
\bibitem{LongPRL2024}
Y. Long, Z. Wang, C. Zhang, H. Xue, Y. X. Zhao, and B, Zhang, Non-Abelian Braiding of Topological Edge Bands, Phys. Rev. Lett. \textbf{132}, 236401 (2024).
\bibitem{Funpj2022}
B. Fu, J.-Y. Zou, Z.-A. Hu, H.-W. Wang, and S.-Q. Shen, Quantum anomalous semimetals, npj Quantum Mater. \textbf{7}, 94 (2022).

\bibitem{HuPRL2021}
H. Hu, and E. Zhao, Knots and Non-Hermitian Bloch Bands, Phys. Rev. Lett. \textbf{126}, 010401 (2021).
\bibitem{HuPRR2022}
H. Hu, S. Sun, and S. Chen, Knot topology of exceptional point and non-Hermitian no-go theorem, Phys. Rev. Research \textbf{4}, L022064 (2022).
\bibitem{GuoPRL2023}
C.-X. Guo, S. Chen, K. Ding, and H. Hu, Exceptional Non-Abelian Topology in Multiband Non-Hermitian Systems, Phys. Rev. Lett. \textbf{130}, 157201 (2023).
\bibitem{WangNanotechnology2025}
Y. Wang, Y. Wu, X. Ye, C.-K. Duan, Y. Wang, H. Hu, X. Rong, and J. Du, Non-Hermitian non-Abelian topological transition in the $S=1$ electron spin system of a nitrogen vacancy centre in diamond, Nat. Nanotechnol. \textbf{20}, 873–880 (2025).
\bibitem{TangNSR2022}
W. Tang, K. Ding, and G. Ma, Experimental realization of non-Abelian permutations in a three-state non-Hermitian system, Natl. Sci. Rev. \textbf{9}, (2022).
\bibitem{BaoPRB2024}
L. Bao, A. Shi, Y. Peng, P. Peng, J. Ning, Z. Wang, C. Peng, and J. Liu, Non-Abelian permutations meet $D_4$ group in a four-state non-Hermitian system, Phys. Rev. B. \textbf{110}, L020104 (2024).



\bibitem{ZhangNP2022}
X.-L. Zhang, F. Yu, Z.-G. Chen, Z.-N. Tian, Q.-D. Chen,
H.-B. Sun, and G. Ma, Non-Abelian braiding on photonic chips, Nat. Phys. \textbf{16}, 390–395 (2022).


\end{thebibliography}
\end{document}